\documentclass[letterpaper]{jpconf}
\usepackage{graphicx}
\begin{document}
\title{Structure Function Measurements at HERA}

\author{Alexey Petrukhin}

\address{DESY, 22607 Notkestr. 85, Hamburg, Germany}
\address{ITEP, 117218 Bolshaya Cheremushkinskaya 25, Moscow, Russia}

\ead{petr@mail.desy.de}

\begin{abstract}
Recent structure function results from the H1 and ZEUS Collaborations are presented.
The data cover a wide
kinematic range of squared four-momentum transfers $Q^2$, from 0.2\,GeV$^2$ to
30000\,GeV$^2$, and Bjorken $x$ between $\sim$$5*10^{-6}$ and 0.65. Data from both
experiments have been combined, leading to significantly reduced
experimental uncertainties. The combined measurements are analysed using
a NLO QCD fit, and a set of parton density functions, HERAPDF1.0, is
extracted.
New direct measurements of the structure function $F_L$, making use of
dedicated low energy runs of the HERA machine, are also presented.

\end{abstract}

\section{Measurements of the structure function $F_2$}
A new measurement [1] of deep inelastic lepton-nucleon scattering (DIS) is based on data collected by the H1 collaboration in the year 2000 with positrons of energy $E_e$=27.6\,GeV and protons of energy $E_p$=920\,GeV, corresponding to a centre-of-mass energy $\sqrt{s}$=319\,GeV. The measurement is performed in the kinematic region of 12\,GeV$^2 \leq Q^2 \leq$ 150\,GeV$^2$ and of $10^{-4} \leq x \leq 0.1$. The luminosity amounts to 22 pb$^{-1}$. This measurement is combined with similar H1 data taken in 1996/97 at $E_p$=820\,GeV [2]. The combined data represent the most precise measurement in presented kinematic domain with typical total uncertainties in the range of 1.3-2\%.   
%The measurement corresponds to a wide range of inelasticity $y$, from 0.005 to 0.6. 
The data are used to determine the structure function $F_2(x,Q^2)$, which is observed to rise continuously towards low $x$ at fixed $Q^2$. A NLO QCD analysis is performed to obtain a new set of parton distribution functions H1PDF2009 [1] from the inclusive DIS cross section measurements presented here as well as from previously published H1 measurements at low [3] and high [2] $Q^2$. The data and the NLO QCD fit from H1 data alone are shown in Figure 1.    

\section{Combined H1 and ZEUS measurements}

A combination [4] is presented of the inclusive DIS cross sections measured by the
H1 and ZEUS Collaborations in neutral current unpolarised $ep$ scattering at
HERA during the period 1994-2000. The luminosity amounts to 240 pb$^{-1}$. The data cover a several orders of magnitude in $Q^2$, and in Bjorken $x$. The combination method used
takes the correlations of systematic uncertainties into account, resulting in an improved
accuracy. 
The input data from H1 and ZEUS are consistent with each other
at $\chi^2/$ndf = 636.5/656. The total uncertainty of the combined data set reaches 1\%
in the best measured region, 20\,GeV$^2 < Q^2 < $ 100\,GeV$^2$. Figure 2 shows the combined HERA results: scaling violations predicted by the theory of QCD and the HERAPDF1.0 fit which will be discussed in the next section.

\begin{figure}[h]
\begin{center}
\centerline{\includegraphics[width=.55\columnwidth]{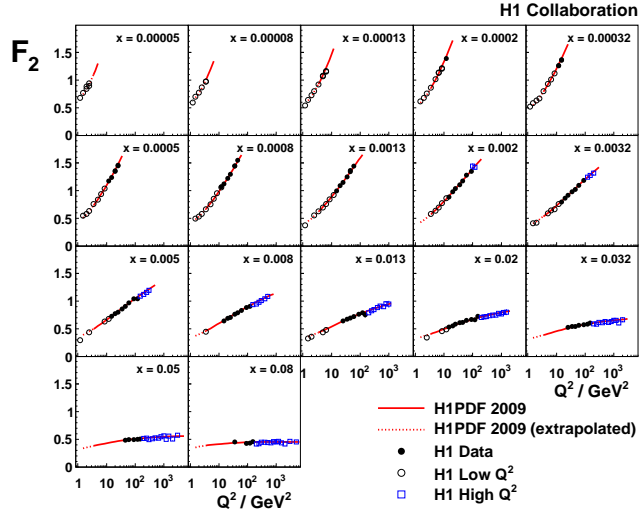}}
\end{center}
\caption{\label{label}Measurements of the structure function $F_2$ as a function of $Q^2$ at various values of $x$. The new data (closed circles) are complemented by the previously published data at low $Q^2$ (open circles) [3] and high $Q^2$ (open boxes) [2]. The error bars represent the total measurement uncertainties. The solid curve represents the NLO QCD fit to H1 data alone for $Q^2\geq3.5\,$GeV$^2$, which is also shown extrapolated down to $Q^2=1.5\,$GeV$^2$.}
\end{figure}

\begin{figure}[h]
  \begin{minipage}[c]{0.5\textwidth}
    \includegraphics[width=1.\columnwidth]{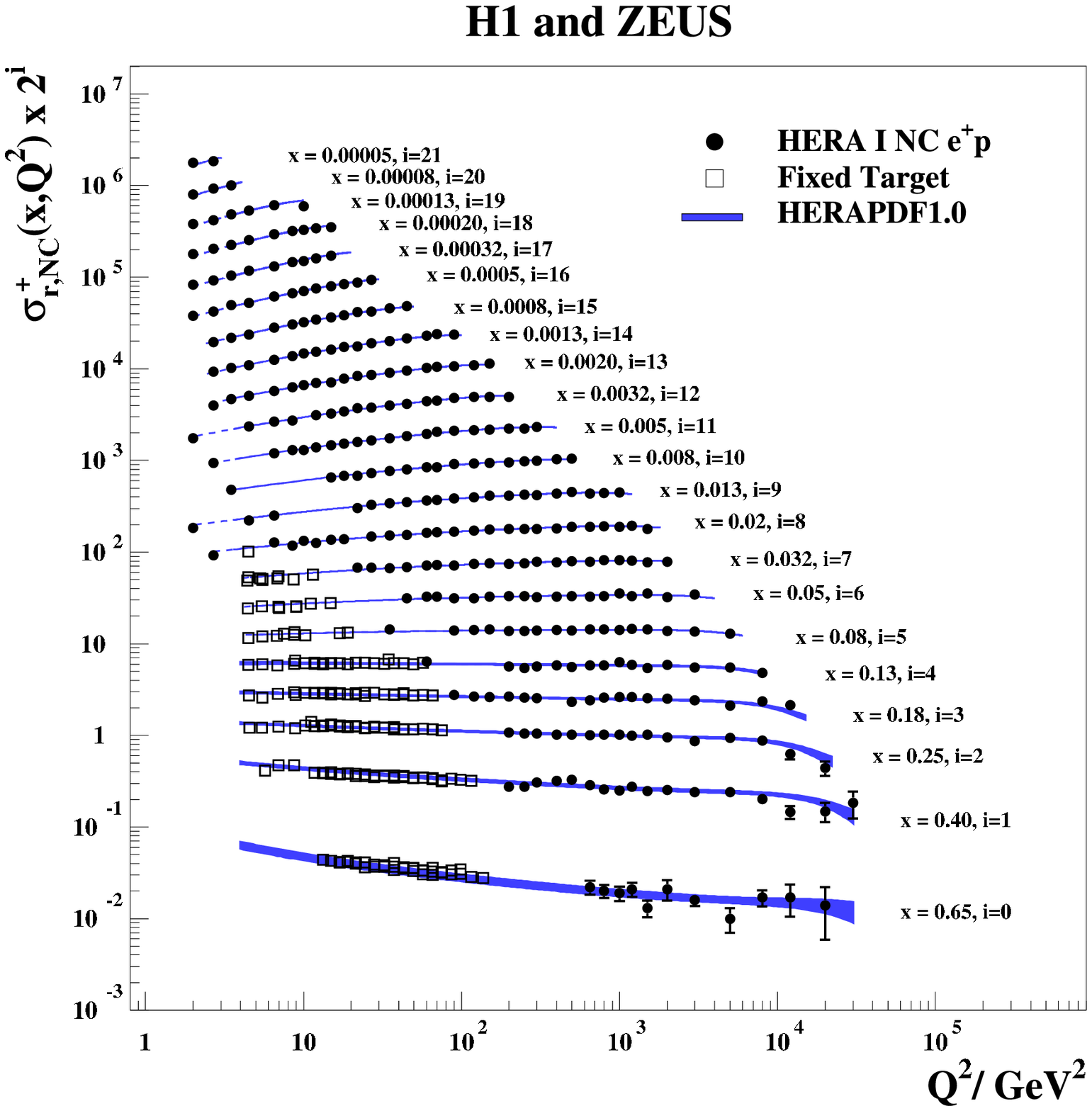}
  \end{minipage}
  \begin{minipage}[c]{0.5\textwidth}
    \includegraphics[width=1.\columnwidth]{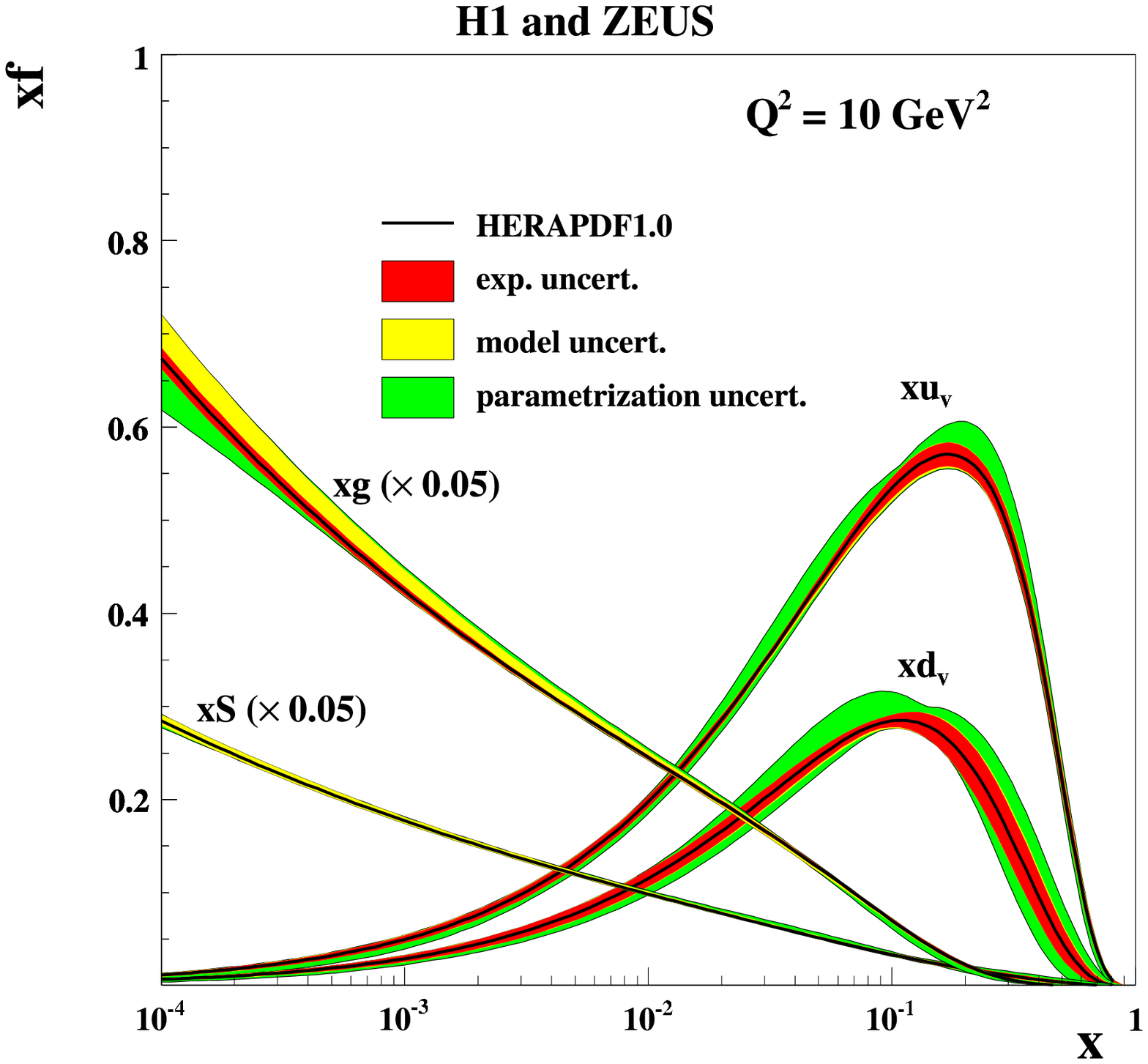}
  \end{minipage}
\caption{\label{fig:contr}
Left: HERA combined neutral current reduced cross section [4] and fixed-target data compared to the HERAPDF1.0 fit. The bands represent the total uncertainty of the fit. Right: the parton distribution functions from the HERAPDF1.0 at $Q^2$ = 10\,GeV$^2$. The gluon and sea distributions are scaled
down by a factor 20. The experimental, model and parametrisation uncertainties are shown
separately (see [4]).
}
\end{figure}

\section{QCD analysis of the combined data}

The combined data set on inclusive cross sections is used as the sole
input for a next-to-leading order QCD analysis which determines
a new set of parton distributions HERAPDF1.0 with small experimental uncertainties. This
set includes an estimate of the model and parametrisation uncertainties of the fit result as explained in [4]. The HERAPDF1.0 fit results are shown in Figure 2. Due to the precision of the combined data set, the parametrisation HERAPDF1.0 has total uncertainties at the level of a
few percent at low $x$.

\section{Measurements of the structure function\,$F_L$}

Figure 3 shows the first measurements of the the structure function\,$F_L$ performed by the H1 [5] and ZEUS [6] Collaborations. The measurement of $F_L$ requires several sets of DIS cross sections at fixed $x$ and $Q^2$ but at different values of inelasticity $y$. This was achieved at HERA by variations of the proton beam energy whilst keeping the lepton beam energy fixed. The current measurements are
based on inclusive deep inelastic $e^+p$ scattering cross section measurements with a positron beam energy of 27.5\,GeV and proton beam energies of 920, 575 and 460\,GeV. Employing the energy dependence of the cross section, $F_L$ is measured in the range of $12 \leq Q^2
 \leq 130$\,GeV$^2$ and low Bjorken $x$ of $0.00024 \leq x \leq 0.007$. The $F_L$ values agree with higher order QCD calculations based on
 parton densities obtained using cross section data previously measured at HERA.

\begin{figure}[h]
  \begin{minipage}[c]{0.5\textwidth}
    \includegraphics[width=1.\columnwidth]{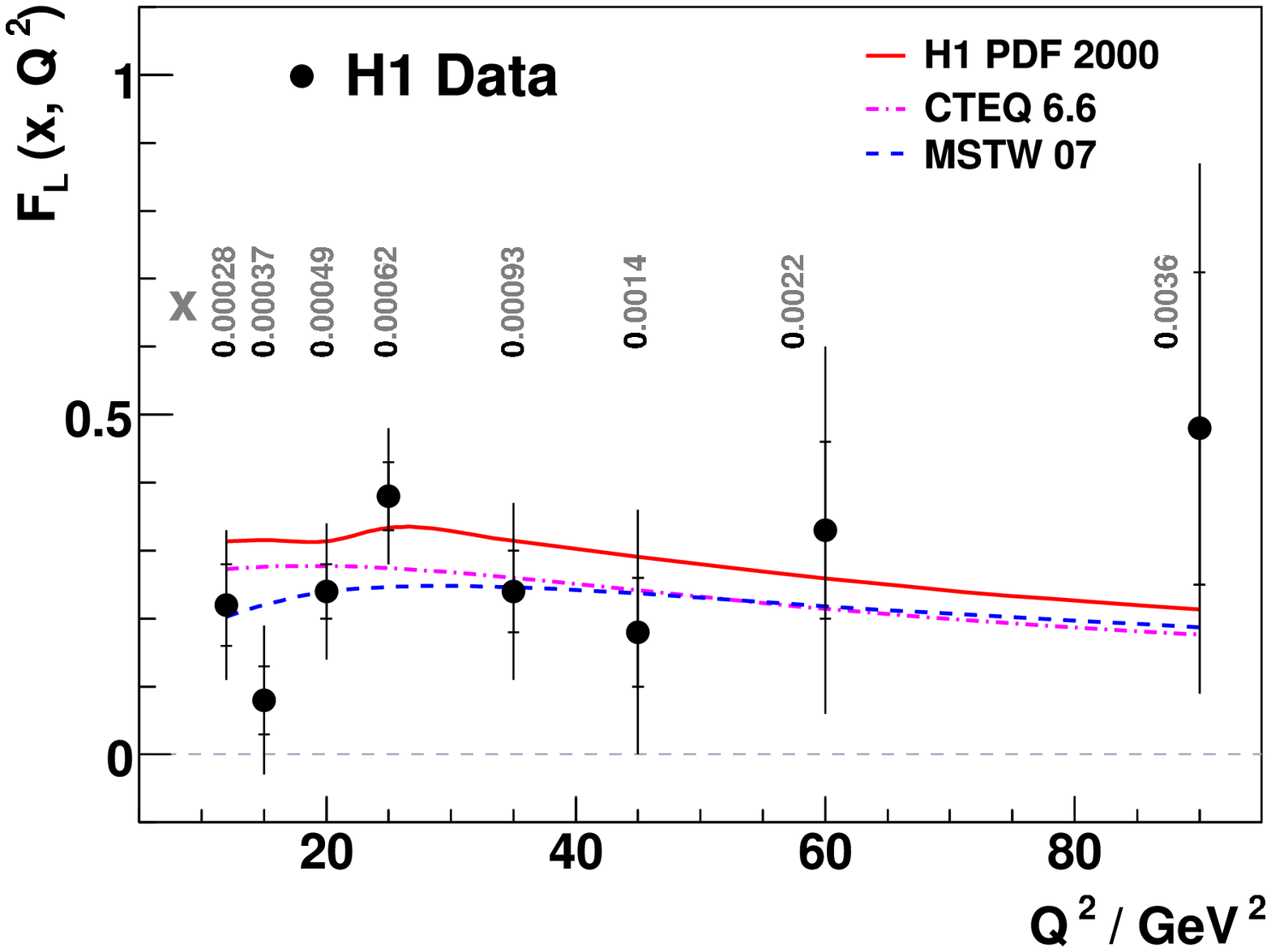}
  \end{minipage}
  \begin{minipage}[c]{0.5\textwidth}
    \includegraphics[width=1.\columnwidth]{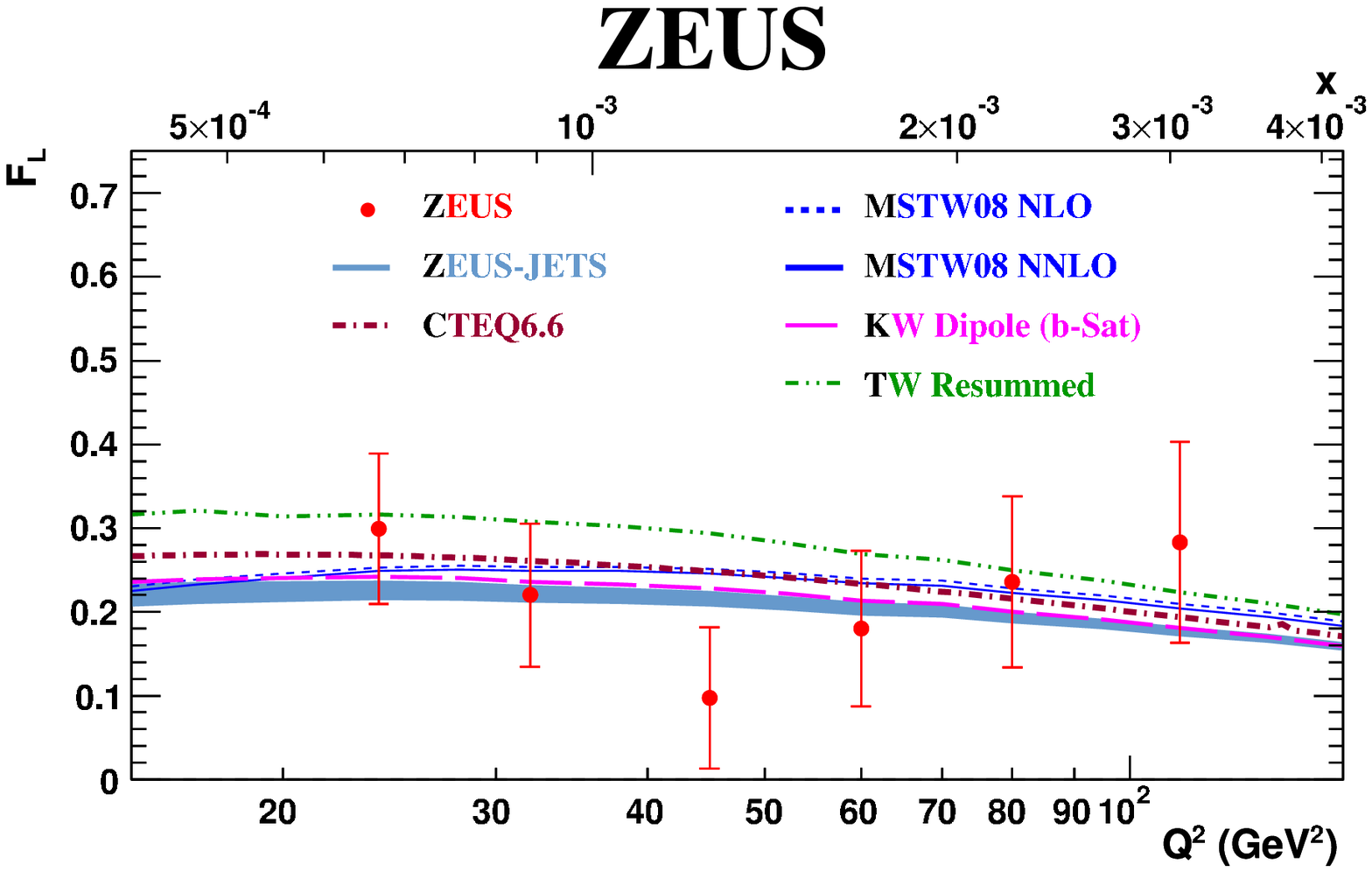}
  \end{minipage}
\caption{\label{fig:contr}
$F_L$ from the H1 data\,[5] (left) and ZEUS data\,[6] (right) compared to the different theoretical predictions. The full error bars
include the statistical and systematic uncertainties added in quadrature.
}
\end{figure}

Figure 4 shows a new preliminary H1 measurement of the structure function $F_L$ [7]. The measurements of $F_L$ use different parts of the H1 detector covering when combined a wide range of squared four-momentum transfers $2.5 \leq Q^2 \leq 800~$GeV$^2$ and Bjorken $x$ between 0.00005 and 0.035. The data are compared with higher order QCD prediction H1PDF2009 [1]. The measurements are in a good agreement with H1PDF2009 fit except the lowest $Q^2$ region.

\begin{figure}[h]
\begin{center}
\centerline{\includegraphics[width=.6\columnwidth]{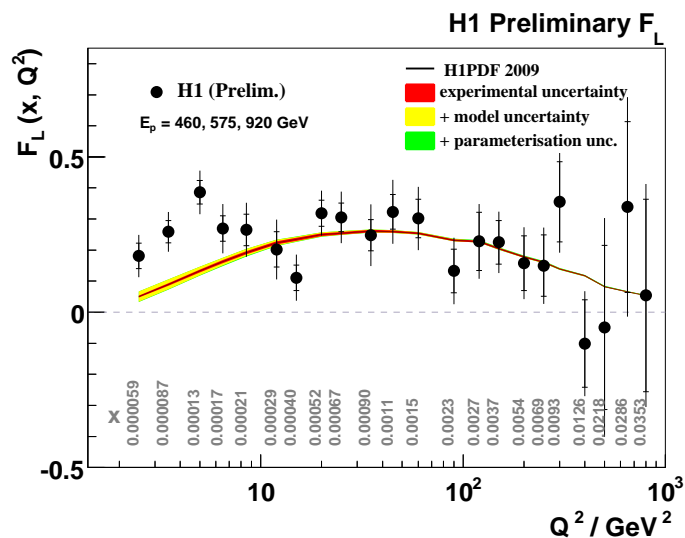}}
\end{center}
\caption{\label{label}$F_L$ from the H1 data\,[7] in extended kinematic range of squared four-momentum transfers $2.5 \leq Q^2 \leq 800~$GeV$^2$ and $0.00005 \leq x \leq 0.035$. The data are compared to NLO QCD fit H1PDF2009. The full error bars include the statistical and systematic uncertainties added in quadrature.}
\end{figure}

\end{document}